\documentclass[reprint,aps,pra,superscriptaddress,twocolumn]{revtex4-1}

\usepackage{subfig}
\usepackage{amsmath}
\usepackage{graphicx}
\begin{document}


\title{A set of $4d-3$ observables to determine any pure qudit state}



\author{Quimey Pears Stefano}
\email[]{quimeyps@df.uba.ar}
\affiliation{Universidad de Buenos Aires, Facultad de Ciencias Exactas y Naturales, Departamento de F\'isica, Buenos Aires, Argentina.}
\affiliation{Consejo Nacional de Investigaciones Cient\'ificas y T\'ecnicas, Buenos Aires, Argentina.}

\author{Lorena Reb\'on}
\affiliation{Departamento de F\'isica, IFLP-CONICET, Universidad Nacional de La Plata, C.C. 67, 1900 La Plata, Argentina.}

\author{Silvia Ledesma}
\affiliation{Universidad de Buenos Aires,
  Facultad de Ciencias Exactas y Naturales, Departamento de F\'isica,
  Buenos Aires, Argentina.}
\affiliation{Consejo Nacional de
  Investigaciones Cient\'ificas y T\'ecnicas, Buenos Aires,
  Argentina.}
\author{Claudio Iemmi}
\affiliation{Universidad de
  Buenos Aires, Facultad de Ciencias Exactas y Naturales, Departamento
  de F\'isica, Buenos Aires, Argentina.}
\affiliation{Consejo
  Nacional de Investigaciones Cient\'ificas y T\'ecnicas, Buenos
  Aires, Argentina.}

\date{\today}

\begin{abstract}
We present a tomographic method which requires only $4d-3$ measurement
outcomes to reconstruct \emph{any} pure quantum state of arbitrary
dimension $d$. Using the proposed scheme we have experimentally
reconstructed a large number of pure states of dimension $d=7$,
obtaining a mean fidelity of $0.94$. Moreover, we performed numerical
simulations of the reconstruction process, verifying the feasibility
of the method for higher dimensions.  In addition, the \emph{a priori}
assumption of purity can be certified within the same set of
measurements, what represents an improvement with respect to other
similar methods and contributes to answer the question of how many
observables are needed to uniquely determine any pure state.

\end{abstract}

\maketitle
\makeatletter
\newcommand*\rel@kern[1]{\kern#1\dimexpr\macc@kerna}
\newcommand*\widebar[1]{%
  \begingroup
  \def\mathaccent##1##2{%
    \rel@kern{0.8}%
    \overline{\rel@kern{-0.8}\macc@nucleus\rel@kern{0.2}}%
    \rel@kern{-0.2}%
  }%
  \macc@depth\@ne
  \let\math@bgroup\@empty \let\math@egroup\macc@set@skewchar
  \mathsurround\z@ \frozen@everymath{\mathgroup\macc@group\relax}%
  \macc@set@skewchar\relax
  \let\mathaccentV\macc@nested@a
  \macc@nested@a\relax111{#1}%
  \endgroup
}
\makeatother
\newcommand{\rev}[1]{{\color{red} #1}}
\newcommand{\unitMuM}[1]{$#1\,\,\mu\mathrm{ m}$}
\newcommand{\unitCM}[1]{$#1\,\mathrm{cm}$}
\newcommand{\unitNM}[1]{$#1\,\mathrm{nm}$}

\newcommand{\figref}[1]{Figure \ref{#1}}
\newcommand{\etal}{\emph{et al.}}

\newcommand{\ket}[1]{\ensuremath{|#1\rangle}}
\newcommand{\bra}[1]{\ensuremath{\langle #1|}}
\newcommand{\braket}[2]{\ensuremath{\langle#1|#2\rangle}}
\newcommand{\ketbra}[2]{\ensuremath{|#1\rangle\!\langle#2|}}

\newcommand{\Projector}[1]{\ensuremath{\mathrm{Proj}\left(#1\right)}}

\newcommand{\PP}[2]{\ensuremath{\mathrm{\hat{P}}_{#1}^{(#2)}}}

\newcommand{\pk}[2][k]{p_{#2}^{(#1)}}
\newcommand{\ck}[1][k]{c_{#1}}

\newcommand{\nk}[2][k]{n_{#2}^{(#1)}}

\newcommand{\nkbar}[2][k]{\widebar{n_{#2}^{(#1)}}}

\newcommand{\gammak}[1][k]{\gamma_{#1}}
\newcommand{\lambdaosc}{\lambda_{\mathrm{dc}}}

\newcommand{\OBJ}{$\mathbf{OBJ}$}

\newcommand{\SFcero}{$\mathbf{SF_0}$}
\newcommand{\SFuno}{$\mathbf{SF_1}$}
\newcommand{\SFdos}{$\mathbf{SF_2}$}
\newcommand{\SFtres}{$\mathbf{SF_3}$}

\newcommand{\BSuno}{$\mathbf{BS1}$}

\newcommand{\Luno}{$\mathbf{L_1}$}
\newcommand{\Ldos}{$\mathbf{L_2}$}
\newcommand{\Ltres}{$\mathbf{L_3}$}
\newcommand{\Lcuatro}{$\mathbf{L_3}$}
\newcommand{\Lc}{$\mathbf{L_c}$}

\newcommand{\Limg}{$\mathbf{L_\mathrm{img}}$}
\newcommand{\Lft}{$\mathbf{L_\mathrm{ft}}$}

\newcommand{\ImagePlaneA}{$\pi'$}
\newcommand{\ImagePlaneB}{$\pi''$}
\newcommand{\ImagePlaneC}{$\pi'''$}

\newcommand{\IA}{$\mathbf{IA}$}
\newcommand{\FA}{$\mathbf{FA}$}

\newcommand{\SLM}{$\mathbf{SLM}$}
\newcommand{\SLMuno}{$\mathbf{SLM_1}$}
\newcommand{\SLMdos}{$\mathbf{SLM_2}$}

Quantum information processing is a field of research in constant
growth since the last decades. It has been crucial both in the
fundamental tests of quantum mechanics as well in the development of a
vast number of technological applications~\cite{Alber2003quantum}. A
fundamental problem that transverse the entire field is that of
determining the unknown state of a quantum system~\cite{Paris2004}. In
this regard, the process of reconstructing a general state of
dimension $d$, known as quantum tomography, consists in finding its
density matrix $\rho$, a positive semidefinite matrix with unit trace
that fully describes the system.
As the number of independent parameters that defines an arbitrary
density matrix $\rho$ is $d^2-1$, typical quantum tomography schemes
require a number of measurement outcomes that increases as
$d^2$~\cite{James2001,Wootters1989,Adamson2010}, leading to
time consuming measurements and post processing, that can become
prohibitive for high dimension
systems~\cite{Kaznady2009,Moroder2012}. Furthermore, several
applications such as quantum key distribution
\cite{Cerf2002,Dada2011,Mower2013,Zhong2015,Mirhosseini2015}, quantum
communication complexity problems \cite{Martinez2018} or fundamental
tests of quantum mechanics \cite{Canas2014,Vertesi2010}, can be
improved by using $d$-level quantum systems of dimension grater than
two (qudits). Therefore, it is of great interest to develop
alternative methods that employ fewer measurement settings.

A significantly reduction in the number of measurements is feasible if
\emph{a priori} information about the unknown state is taken into
account in the tomographic process. For example, in the case of
permutationally invariant states, a set of measurements that scales
polynomially with the number of qubits $n$ ($d=2^n$) is enough to
perform the reconstruction of the matrix $\rho$
\cite{Moroder2012,Schwemmer2014}, instead of applying the habitual
process with exponential growth.  In the case of mixed states of $n$
qudits a reconstruction scheme which employs only \emph{local}
measurements can efficiently give an estimation of the state
\cite{Baumgratz2013}, provided they are well approximated by matrix
product operators. For low-rank states, compressed sensing techniques
allow the reconstruction of $\rho$ from a number of measurement
outcomes of the order of $d(\log d)^2$ by measuring the expectation
values of enough observables chosen at random~\cite{Flammia2012,
  Gross2011}.

Special attention has been paid to the reconstruction of pure states.
It has been shown~\cite{Finkelstein2004} that there exists a
positive-operator-valued measure (POVM) which consists of $2d$
projectors from whose probability outcomes it is possible 
to distinguish \emph{almost} all pure states, except for a set of
measure zero; when all the pure states are included, the number of
required measurement outcomes increases.  In Ref.~\cite{Wang2018},
Wang \textit{et al.} have proven that a set of $4d-3$ fixed projectors
are enough to reconstruct \emph{any} pure qudit. Additionally, they
showed that this number can be reduced to at most $3d-2$ in an
adaptive scheme, where the results of a first measurement of the
particular unknown state in the canonical base ($d$ projections) are
used to choose the remaining projectors.

A highly desirable feature of a tomographic scheme that uses \textit{a
  priori} information, is that it also allows to verify if the initial
assumption about the unknown state is true~\cite{Chen2013}.  In this
regard, Goyeneche \textit{et al.}~\cite{Goyeneche2015} have proposed,
and experimentally tested, an adaptive method that requires a set of
five measurement bases to determine \textit{any} pure qudit in
arbitrary dimension, and additionally certify the assumption of purity
without extra measurement outcomes. In addition, Carmelli \textit{et
  al.} have constructed in~\cite{Carmeli2016} a set of five fixed
bases for the tomography of any pure state.

Recently~\cite{PearsStefano2017}, we have reported a quantum
tomography method to reconstruct \emph{any} pure photonic spatial
qudit of arbitrary dimension from a set of only $4d$ measurement
outcomes, in case of a fixed measurement settings, or $4d-3$ when the
procedure is performed in an adaptive way.  In that work, the qudits
were codified in the discretized transverse momentum of single photons
(\emph{slit} states) \cite{Neves2004}.  The proposed experimental
setup exploited the spatial nature of the codification to set a
Mach-Zehnder interferometer in the state-estimation stage. This
interferometer is used to implement the \emph{three-step} Phase Shifting
Interferometry technique (PSI)~\cite{Creath1988}, which is ideally
suited to find the phase of a light wavefront. As the method allows
to reconstruct the whole wavefront from the interference pattern, the
$4d$ measurement outcomes can be obtained in parallel by means of an
image detector, reducing the measurement settings to only four,
independently of the dimension $d$ of the quantum
system~\cite{PearsStefano2017}.  This measurement scheme is extremely
efficient, and work properly for spatial qudits where it is
straightforward to parallelize the measurements. However, still
missing an experimental implementation of such tomographic method for
more general encoding systems.

In this Letter we generalize that result to reconstruct \emph{any}
unknown pure qudit of arbitrary dimension $d$ from the outcome
frequencies of $4d-3$ projectors, which do not depend on the physical
representation of the state. While this amount of projectors
represents approximately $d$ more measurement outcomes than reported
in Ref.~\cite{Wang2018} for an adaptive scheme, our method allows us
to certify the assumption of purity. Conversely, our method requires
in the order of $d$ measurements less than the method reported in
Ref.~\cite{Carmeli2016}, but at the expense of using adaptive
measurements. Then, our scheme close the gap in the number of
measurement between both kind of methods suggesting that, in order to
have a fixed set of observables, the number of measurements has to be
increase to $5d$. Instead of that, if the number of measurements is
expected to be reduced to $\sim 3d$, the possibility of discriminate
pure from mixed states is lost.

Let us to start by describing the reconstruction process scheme. Any
pure quantum state of dimension $d$ can be expressed as
\begin{eqnarray}\label{qudit}
|\Psi\rangle =\sum_{k=0}^{d-1} c_k |k\rangle,
\end{eqnarray}
where the $c_{k}~'s$ are complex coefficients and $|k\rangle$ denotes
one of the $d$ states of some base of the Hilbert space, in particular
the canonical base. These complex coefficients can be explicitly
written as $c_k=|c_k| e^{i \varphi_k}$, being $\varphi_k$ a real
number. For each element \ket{k} of the canonical base such that
$k=0,2,3,...,d-1$, we can define a set of $d$-dimensional states
\begin{equation}
  |\Psi_{\ell}^{(k)}\rangle =\left(\ket{0} +
    e^{i\pi/2\times\left(\ell-1/2 \right)} \ket{k}\right)
  /\sqrt{2},~\ \ \ \ \ \ \ \ell=1,2,3 \label{eq:projectors}
\end{equation}
from which $3d-3$ projectors
$\hat{P}_{\ell}^{(k)}=|\Psi_{\ell}^{(k)}\rangle\langle\Psi_{\ell}^{(k)}\vert$
are obtained. The outcome probabilities of these projectors
$p_{\ell}^{(k)} =
\bra{\Psi}\hat{P}_{\ell}^{(k)}\ket{\Psi}=\vert\langle\Psi_{\ell}^{(k)}\ket{\Psi}\vert^2$,
i. e., the probabilities of measuring every state
$|\Psi_{\ell}^{(k)}\rangle$ in the state $|\Psi\rangle$, are related
to the coefficients $c_{k}~'s$ by the set of $d-1$ equations
\begin{equation}
  \sqrt{2}c_0 c_{k}^*=\left(p_{1}^{(k)}-p_{2}^{(k)}\right) +
  i \left(p_{3}^{(k)}-p_{2}^{(k)}\right),
\label{eq:phase-eq}
\end{equation}
where, without loss of generality, we have define $c_0$ as real. Thus,
with the knowledge of $c_0\equiv +\sqrt{p_0}$, which is obtained from
the probability $\vert\langle0\ket{\Psi}\vert^2=p_0$, the remaining
$c_k$ are determined.  These $3d-2$ measurement
outcomes are sufficient for determining a generic pure state, with the
exception of a set of measure zero, corresponding to those with a null
value of $c_0$. This limitation can be overcome with the addition of a
previous measurement of the unknown state onto the canonical base,
from which we redefine the element $\ket{0}$ to satisfy the condition
$|c_0|=\underset{k}{Max} \lbrace|c_k|\rbrace$. 
This leads to require determining the average values of $4d-3$
observables to completely solve the the pure-states reconstruction
problem.

A remarkable feature of the method is that the same set of
\mbox{$4d-3$} measurement outcomes allows to verify the \emph{a
  priori} assumption of purity. Following a similar argument to the
one in Ref.~\cite{Goyeneche2015}, lets us consider a quantum state of
dimension $d$ represented by the density matrix $\rho$.  If the state
is pure, there is an unitary matrix $U$ that diagonalizes $\rho$ such
that $\rho= U \tilde{\rho} U^\dagger$, with $\tilde{\rho}_{kl}=1$ for
$k=l=1$ and $0$ in any other case.  Hence, it is easy to see that:
\begin{eqnarray}\label{pure}
\rho~~\mathrm{is~pure} \Leftrightarrow |\rho_{k,l}|^2=|\rho_{k,k}||\rho_{l,l}|~,~~\forall k,l=0,\ldots,d-1.
\end{eqnarray} 
Additionally, being $\rho$ a Hermitian matrix it can be written as
$\rho = A A^\dagger$ for some operator $A$. As a consequence, its
elements can be expressed as
$\rho_{k,l} = \langle\textbf{v}_k , \textbf{v}_l\rangle$, where
$\{\textbf{v}_k\}_{k=0}^{d-1}$ are the row vectors of $A$, and
$\langle~,~\rangle$ indicates the inner product. According to the
Cauchy-Schwarz statement Eq.~(\ref{pure}) holds iff $\textbf{v}_k$ and
$\textbf{v}_l$ are linearly dependent, and therefore the state will be
pure iff the rows of $A$ are \textit{all} parallel. Finally, to
validate the purity assumption is enough to verify the $d-1$ equations
 \begin{equation}
   |\rho_{0,k}|^2=|\rho_{0,0}||\rho_{k,k}|~,~~k=1,\ldots,d-1,
   \label{eq:cauchy-schwarz}
 \end{equation} 
 which are hold iff every $\textbf{v}_k$ with $k=1,\ldots,d-1$ is
 parallel to $\textbf{v}_0$, and by transitivity, they are all
 parallel to each other. Physically, $|\rho_{0,k}|$ is the visibility
 of the two slits interference pattern, and in the context of the
 present tomographic scheme, it can be calculated as
 $\sqrt{\frac{\left(\pk{1}-\pk{2}\right)^2
     +\left(\pk{2}-\pk{3}\right)^2}{2}}$. Besides, the right-hand side
 of Eq.~(\ref{eq:cauchy-schwarz}) is the product of the relative
 intensities of these two slits in the encoding state, which are
 readily evaluated from a measurement in the canonical base.  It is
 worth to remember the close relationship between the set of
 projectors $\hat{P}_{\ell}^{(k)}$ and the method of three step
 PSI. The index $\ell$ in Eq.~(\ref{eq:projectors}) is related to an
 increment in steps of $\pi/2$ of the phase difference between
 $\ket{k}$ and the reference state $\ket{0}$. For every $k$, these
 three projectors correspond to the interference between the object
 beam and the reference beam for phase differences of $\pi/2$, $\pi$
 and $\pi/2$. In fact, it can be shown that Eq.  (\ref{eq:phase-eq})
 leads to $\varphi_k = \arctan\frac{\pk{3}-\pk{2}}{\pk{1}-\pk{2}},$
 which is the explicit solution for the phases $\varphi_k$ in the PSI
 method \cite{Creath1988}.

\begin{figure}[htbp]
\centering
\includegraphics[width=.92\linewidth]{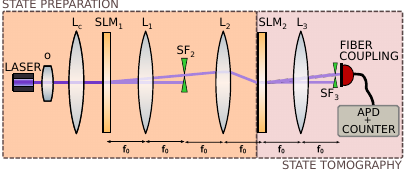}%
\caption{Experimental setup. The light source is a
  405nm cw laser diode, attenuated down to the single photon level.\\
  \textbf{Ls}: convergent lenses; \textbf{SLMs}: pure phase spatial
  light modulators; \textbf{SFs}: spatial filters. The detection in
  the centre of the interference pattern is performed with a
  fiber-coupled \textbf{APD}.  }
\label{fig:exp-setup}
\end{figure}

To experimentally test the tomographic method in the context of
projective measurements we have used the set-up schematically
depicted in Figure \ref{fig:exp-setup}.
This setup can be divided in two modules, the first one is employed
for the state preparation (SP) and the second one is used to perform the
state tomography (ST). Let us start by describing the SP part that is
basically a 4-f optical processor. The light source is a 405nm laser
diode, attenuated to the single photon level. The beam is expanded by
a microscope objective (O), spatially filtered ($\mathrm{SF}_1$) and
then collimated by the $\mathrm{L}_c$ lens in such a way that onto the
first Spatial Light Modulator (SLM), placed at the front focal plane
of lens $\mathrm{L}_1$, a plane wave impinges with almost constant
intensity distribution onto the region of interest (ROI). The SLM
consists of a Sony liquid crystal television panel model LCX012BL that
in combination with polarizers and wave plates, that provide the
adequate state of polarization of light, allows a full phase
modulation of the incident wavefront for the operating
wavelength~\cite{Marquez2001}. By using the method described in the
Ref.~\cite{Solis-Prosser2013} it is possible to obtain at the back
focal plane of lens $\mathrm{L}_2$ the complex amplitudes $c_k$ of an
arbitrary slit state.  Briefly, in this method each slit is
represented by a phase grating. The absolute amplitude of each slit is
controlled by means of the phase modulation of the grating as the
diffraction efficiency depends on it. To encode this information we
chose the first diffracted order, which is selected by
($\mathrm{SF}_2$).  The argument of $c_k$ is obtained by adding an
adequate uniform phase on the grating modulation.  The ST is performed
by using ($\mathrm{SLM}_2$) to provide the necessary phase modulation
to encode the projector state by applying the same method used in the
SP part. This second SLM is placed at the front focal plane of lens
$\mathrm{L}_3$, so, after filtering the first diffracted order by
means of $\mathrm{SF}_3$, the exact Fourier transform of the projected
spatial qudit is obtained at the detector plane. The light
distribution corresponds to the interference pattern projection
between the prepared state and the selected projector state. The
single photon count rate in the centre of the interference pattern is
proportional to the probability of projection of the two
states~\cite{Lima2011}. The detector module is a fiber-coupled
avalanche photodiode (APD) photon counting module Perkin Elmer
SPCM-AQRH-13-FC. The core aperture of the fiber ($\Phi=$
\unitMuM{62.5}) acts as a pupil that selects the centre of the
interference pattern.
\begin{figure}[htbp]
  \centering
  \includegraphics[width=0.385\textwidth]{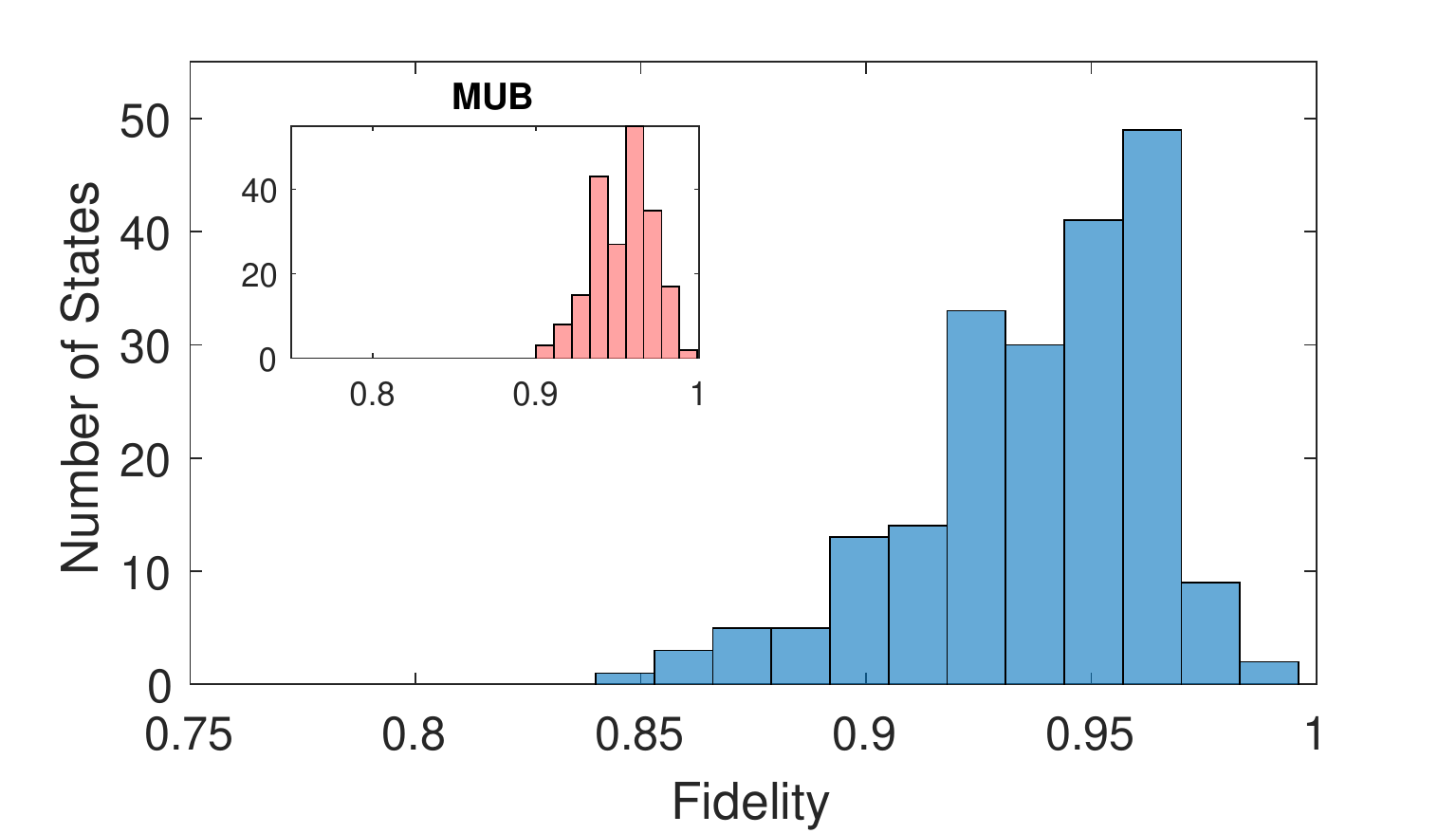}
  \caption{Histogram of the experimental reconstruction fidelity for
    200 random states of dimension $d = 7$. The mean fidelity is
    $\bar{F} = 0.94$ and the standard deviation is
    $\sigma_F=0.03$. The inset shows the experimental reconstruction
    fidelity, for the same 200 states using the MUB method. The mean
    fidelity in this case is $\bar{F_\mathrm{mub}} = 0.95$, and the
    standard deviation is $\sigma_{F_\mathrm{mub}} = 0.03$. For this
    dimension our method only requires 25 projective measurements,
    while the MUB method requires 56.}
\label{fig:histogram-fidelity}
\end{figure}

We performed the reconstruction of a large number of pure states for
dimension $d = 7$. As a figure of merit of the reconstruction process
we calculated the fidelity
$F \equiv
\mathrm{Tr}\left(\sqrt{\sqrt{\varrho}\rho\sqrt{\varrho}}\right)$
between the state intended to be prepared, $\varrho$, and the density
matrix of the reconstructed state, $\rho$ \cite{Jozsa1994}.
Fig. \ref{fig:histogram-fidelity} shows an histogram of the
reconstruction fidelity for 200 random pure states of dimension
$d = 7$. The mean fidelity is 0.94, with a standard deviation of
0.03. For comparison, we have also reconstructed the same set of
states by means of a standard quantum tomography method using mutually
unbiased bases (MUBs) \cite{Wootters1989}. In this case the mean
fidelity is 0.95, with a standard deviation of 0.03. However, in our
method only 25 projective measurements are necessary instead of the 56
used for implementing the MUBs method.
\begin{figure}[htbp]
  \centering
  \includegraphics[width=0.385\textwidth]{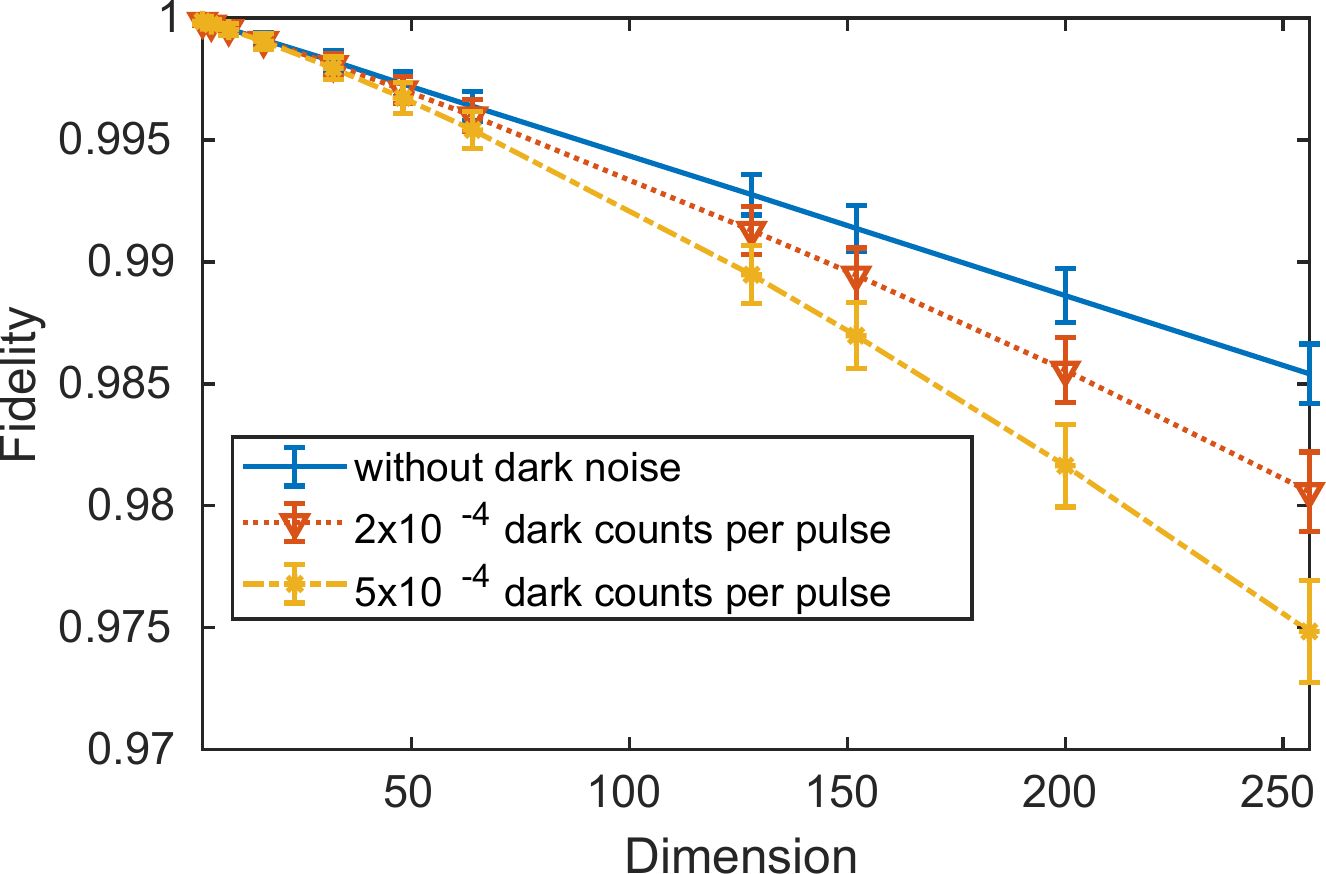}
  \caption{Simulated reconstruction fidelity as a function of the
  dimension. For each dimension, 2000 random states are
  reconstructed. The points represent the mean fidelity value, while
  the error bars are the standard deviation. Each curve corresponds to
  one simulation with a different average dark noise per pulse.}
\label{fig:fidelity-vs-dimension}
\end{figure}

The feasibility of this technique was also tested numerically for
several dimensions of the quantum system and different levels of
experimental noise. To represent a realistic experimental
implementation we assumed a pulsed attenuated laser as a source of
weak coherent states, with a low mean number of photons per pulse
$\mu$. We also include in our simulation a parameter $\lambdaosc$
representing the mean dark counts per pulse caused by self triggering
effects in the APD.  As both, the photon emission process and the dark
counts effect, are assumed to have a Poissonian statistic, the
probability of detecting a photon after projecting $|\Psi\rangle$ onto
the state $|\Psi_{\ell}^{(k)}\rangle$, within a given light pulse, is
\begin{equation}
  \mathrm{Prob}\left(1\ \mathrm{count}; \hat{P}_{\ell}^{(k)}\right) =
  1-\exp\left(-\mu \pk{\ell}-   \lambdaosc\right).
  \label{eq:prob-one-photon}
\end{equation}
If a fixed number of pulses $N$ is considered, the distribution for
the detected counts is a Binomial distribution where the success
probability is given by Eq. (\ref{eq:prob-one-photon}), and the mean
number of counts in the state $|\Psi_{\ell}^{(k)}\rangle$ results
\begin{equation}
  \nkbar{\ell} = N\left\{ 1-\exp\left(-\mu \pk{\ell} -
      \lambdaosc\right)\right\} . 
  \label{eq:N-photon-counts}
\end{equation}
For small $\mu$ Eq. (\ref{eq:N-photon-counts}) is reduced to
$ \nkbar{\ell} \approx N \mu \pk{\ell} + N \lambdaosc,$ which makes
straightforward to evaluate the state-reconstruction equation
(Eq.~(\ref{eq:phase-eq})). Moreover, in our simulations we have used
Eq.~(\ref{eq:N-photon-counts}) to yield a closer estimation of the
projection probabilities for every value of the parameters $\mu$ and
$\lambdaosc$.

In order to perform the numerical simulations we selected $\mu = 0.18$
what results in $84\%$ of empty pulses and about $2\%$ of pulses with
more than one photon, and a value $N = 5\times 10^4$ to guarantee good
statistics. Figure \ref{fig:fidelity-vs-dimension} shows the mean
fidelity of the simulated results as a function of the state dimension
for three levels of noise: without dark counts per pulse (solid line),
$\lambdaosc = 2\times 10^{-4}$ (dotted line) and
$\lambdaosc = 5\times 10^{-4}$ (dashed line) counts per pulse
respectively. These last values are consistent with typical
experimental situations, i. e., a pulse duration of the order of
microseconds and an APD operating in the range of
$100 \mathrm{counts/sec}$. The error bar represents the standard
deviation on the totality of states chosen randomly in the Hilbert
space of dimension $d$.  The fidelity values obtained from these
simulations show that the proposed reconstruction method can be
implemented in real high dimension systems.

\begin{figure}[htbp]
  \centering
  \includegraphics[width=0.385\textwidth]{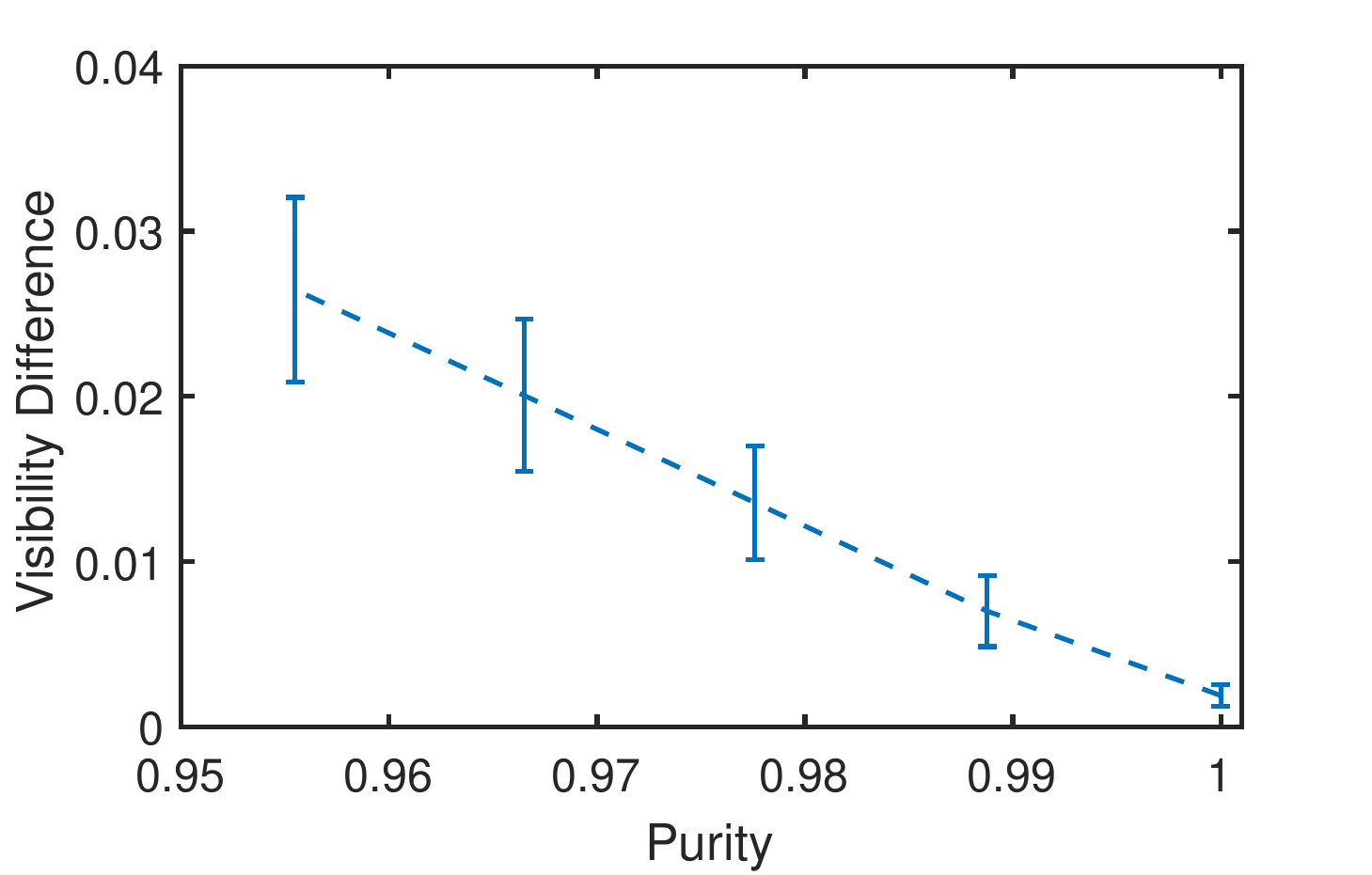}
  \caption{Visibility difference as a function of the degree of purity
    of the state. The curve is averaged over 2000 random states of
    dimension $d=8$. The points represent the average over states,
    while the error bars represent the standard deviation.}
\label{fig:visibility-vs-purity}
\end{figure}

As the Eq.~(\ref{eq:cauchy-schwarz}) only holds for pure states, the
difference between both sides of the equality, i.e., the visibility
difference between a pure quantum state compatible with the given
measurement results and the unknown state that was actually measured,
can be used as a discriminator between pure and mixed states. We
tested this fact by numerically simulating the reconstruction with our
tomographic scheme for random states of dimension $d=8$ affected by
white noise. These states can be written as
\begin{equation}
  \rho_{\mathrm{w}}(|\Psi\rangle, \lambda)
  =(1-\lambda)|\Psi\rangle\langle\Psi| + \lambda \mathbf{1}/d,
  \label{eq:white-noise}\end{equation}
with $\lambda$ a real parameter related to the state purity by means
of
$\mathcal{P}(\rho_{\mathrm{w}})\equiv \mathcal{P}(\lambda)
=\frac{d-1}{d} (\lambda-1)^2+\frac{1}{d}.$ Figure
\ref{fig:visibility-vs-purity} shows the maximum difference between
both size of Eqs.~(\ref{eq:cauchy-schwarz}) as a function of the
purity of the state $\rho_{\mathrm{w}}$, for a fixed value of
$\lambda$. Each point in the curve represents the mean value of such a
difference averaged over 2000 random states $|\Psi\rangle$.  Hence,
given a threshold of purity, this difference can be used to certify
that the state surpass this value.

Summarizing, we have proposed a tomographic method to reconstruct
\emph{any} pure qudit
state 
from only $4d-3$ projective measurement outcomes.  Experimentally, we found
good reconstruction fidelities for states of dimension $d=7$,
comparable with the reconstruction fidelities obtained from the
standard quantum tomography method using MUBs, but with the advantage
of requiring a number of measurement outcomes that scale linearly with
$d$, instead of $d^2$. Furthermore, we have shown through simulations
that our scheme works properly in higher dimension systems, and it is
suitable to distinguishing between pure and mixed states.

\begin{acknowledgments}
  The authors thank D. Goyeneche for helpful discussions.  This work
  was supported by the Agencia Nacional de Promoci\'on de Ciencia y
  T\'ecnica ANPCyT (PICT 2014-2432) and Universidad de Buenos Aires
  (UBACyT 20020170100564BA).
\end{acknowledgments}

\bibliography{biblio_mendeley}

\end{document}